# The Origin of Half-Metallicity in Conjugated Electron System-a Study on Transition Metal Doped Graphyne

Lida Pan,[1] Boqun Song,[1] Jiatao Sun,[1] Lizhi Zhang,[1] Werner Hofer,[2] Shixuan Du,[1] Hong-jun Gao[1]

[1]Institute of Physics, Chinese Academy of Sciences, Beijing 100190, China
[2]Department of Physics, the University of Liverpool, Liverpool L69 3BX, UK
E-mail: sxdu@iphy.ac.cn

**Abstract**

We studied the mechanism of half-metallicity (HM) formation in transition metal doped (TM) conjugated carbon based structures by first-principles electronic structure simulations. It is found that the HM is a rather complex phenomenon, determined by the ligand field splitting of *d*-orbitals of the transition metal (TM) atoms, the exchange splitting and the number of valence electrons. Since most of the conjugated carbon based structures possess ligands with intermediate strength, the ordering of the *d*-orbital splitting is similar in all structures, and the HM properties evolve according to the number of valence electrons. Based on this insight we predict that Cr-, Fe-, Co-doped graphyne will show HM, while Mn- and Ni-doped graphyne will not. By tuning the number of valence electron, we are thus able to control the emergence of HM and control the energy gaps evolving in majority or minority spin channels.

## 1. Introduction

Half-metallic ferromagnets, a system behaving as a metal in one spin channel and a semiconductor in the other, have attracted much attention due to their potential applications in magneto-optics and spintronics. In 1983, half-metallicity (HM) has been discovered in Heusler alloys which are formed by 3*d* transition metal (TM) combining with main group metals.[1] Recently, half-metallic materials based on carbon (e.g. graphite, carbon nanotubes, graphene) have been proposed, thus opening a new avenue of fabricating half-metallic materials.[2-15] Since the electronegativity of carbon is stronger than that of metal elements, valence electrons will not move freely through the lattice as itinerant electrons in an alloy. Therefore, carbon-based HM materials possess unique features distinct from traditional HM alloys. HM in alloys is sensitive to structural features. For example, NiMnSb exhibits HM only in the case of Mn being tetrahedrally positioned.[16] By contrast, HM in carbon based structures is a robust and widespread phenomenon, which is insensitive to crystal symmetries, doping environments or dimensions. Another difference is the concentration of *d*-elements in alloys is much higher than carbon HM materials. In Heusler alloys, the distance between neighboring TM atoms in Heusler Alloy is around 1 Å and *d*-electrons are widely considered itinerant[17]. By contrast, a low concentration of TM doping in carbon is sufficient to give rise to semiconductor-HM transition[2,5]. The large separation removes most of the overlap between *d*-orbitals，thus the conducting mechanism should be distinct from that in alloys.

Numerous works[2-15] have emerged reporting possible carbon HM materials since the successful synthesis of graphene and graphdiyne[18,19]. Most of the work has focused on searching for new HM candidates or tuning HM in specific doped systems[20-24], but failed to search for an underlying principle: the origin of HM in doped carbon based structures. Put succinctly, they failed to answer the question why HM, a rather fragile property in alloys, becomes a robust and widespread property in TM doped-carbon structures. The prevalent HM implies that it originates from some general features of carbon structures, instead of specific structural details.

Here, we are interested in the general HM mechanism at work in TM doped conjugated carbon based structures (TM-carbon). To analyze the physical principles at work we construct a model, which allows us to classify various TM-carbon systems by their principal features like symmetry, coupling strength, or valence electron numbers. Every carbon based TM system has a position in the parameter space established by the model. With the help of first-principle calculations, we are able to specify the position of a concrete system in this parameter space. The symmetry can be different for different TM-carbon systems. However, most TM-carbon systems belong to the same parameter regime—intermediate interaction. This common feature plays an essential role for widespread HM in carbon materials. Considering the energy level splitting due to symmetry and interactions, we suggest a systematic route to generate HM in carbon structure by tuning the valence electron number.

A concrete doped system, TM-graphyne (TM=Cr, Mn, Fe, Co, Ni) is examined by first-principles methods. We find that the conducting properties can be tuned between HM and semi-conducting by varying the doping atom. We expect that the finding is applicable to wide range of TM-carbon systems.

## 2. Methods and model

### 2.1 Basic idea of constructing the model

HM has been discovered in a number of TM-carbon systems,[2-15, 24] which can be seen to consist of two sub-systems: a carbon skeleton and a lattice of TM atoms. The carbon sub-system generally has an inherent gap, being either a semiconductor or a zero-gap semiconductor like graphene; the TM lattice sub-system is insulating in low concentrations. When the two sub-systems are brought together, the combined system can be conducting, semiconducting, or possess HM properties. For instance, if the impurity band appears near the Fermi level, closing the gap by joining the valence and conducting bands of the carbon skeleton, it exhibits conducting properties; if the impurity band is far from Fermi level, the inherent gap of carbon is exposed and the material shows insulating properties. The conducting properties are thus essentially determined by the location of the impurity bands, which depends on the interaction of TM atoms and the carbon skeleton. There are three classes of interactions: TM-carbon interaction, TM internal interaction and TM-TM interactions. In this work, we confine our study to low doping concentration: thus TM-TM interaction is negligible.

### 2.2 Ligand field splitting and exchange splitting

The basic picture of the model is the electron filling of *d*-orbitals, which split under the influence of the surrounding carbon skeleton and TM internal interactions. The splitting caused by the surrounding carbon skeleton here is called ligand splitting. The Hamiltonian of the system is $H = H_0 + V$, where $H_0$ is the Hamiltonian of isolated TM atom and $V$ is the field imposed by the carbon skeleton. $H_0$ has spherical symmetry, while $V$ would typically reduces the symmetry of $H$ and thus reduces the degeneracy of TM-orbitals. The splitting mode of TM orbitals is determined by the symmetry of the carbon skeleton, and is irrelevant to the details of Hamiltonian. The most frequent symmetries in TM-carbon systems are: $D_{3h}$, $C_{3v}$, $C_{6v}$ and $D_{6h}$. For $D_{3h}$, the typical system is a center-doped graphyne (Fig.1a). However, if TM is out-of-plane, the symmetry reduces to $C_{3v}$ (Fig.1c). For $C_{6v}$ symmetry, a typical system is TM doped at the hole of graphene (Fig.1b)[2]. In other cases like carbon nanotubes[3,6], the symmetry is equivalent to doped graphene, if we merely count the influence of the nearest neighbor carbon atoms and neglect the curvature of the nanotube. For $D_{6h}$ symmetry an example is a metal-benzene sandwich (Fig. 1d).[24-26] It can be shown that although there are various symmetries in TM-carbon structures, the representation based on five $d$-orbitals of these frequently-occurring symmetries all contain a single one-dimensional irreducible representation (IR) and two two-dimensional IRs. That is, the $d^5$ orbitals of TM follow a common 2-1-2 splitting mode, $d_{xy}\,d_{x^2-y^2}$, $d_{z^2}$ and $d_{xz},d_{yz}$, under these symmetries. To characterize the splitting mode, we use two parameters: $\Delta$ is the energy difference between $d_{xy}\,d_{x^2-y^2}$ and $d_{z^2}$, $\delta$ is energy difference between $d_{z^2}$ and $d_{xz},d_{yz}$ (Fig. 1).

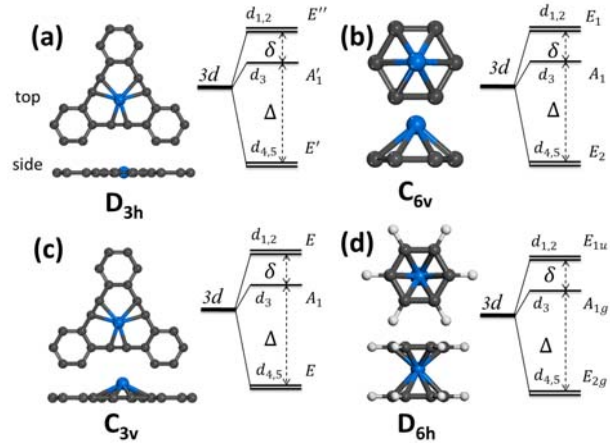

**Figure 1.** Typical symmetries existing in TM-carbon systems and the $d$-orbital splitting mode under these symmetries (only showing the nearest atoms). (a) in-plane doping graphyne with $D_{3h}$ symmetry; (b) doping graphene with $C_{6v}$ symmetry; (c) out-plane doping graphyne with $C_{3v}$ symmetry; (d) metal-benzene sandwich polymers with $D_{6h}$ symmetry. The $d_1$, $d_2$, $d_3$, $d_4$ and $d_5$ refer to the $d$-orbitals of $d_{xz}$, $d_{yz}$, $d_{z^2}$, $d_{xy}$ and $d_{x^2-y^2}$.

In addition to ligand splitting, the splitting caused by TM internal interactions is here called exchange splitting. It originates from the exchange interaction among electrons with the same spin, which leads to the majority spin states having lower energy than the minority spin states. In our model we use a single parameter $J$ to describe the energy shift between the two spin directions.

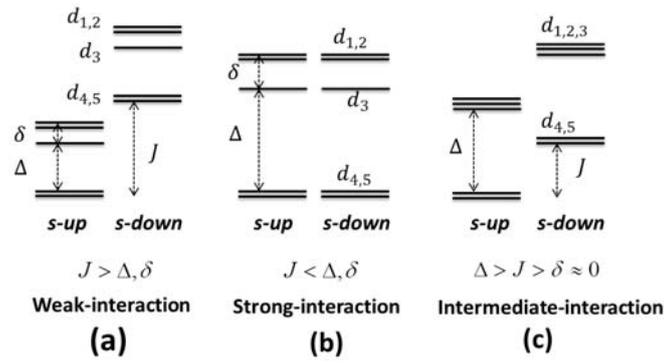

**Figure 2.** The ordering of the *d* orbitals in three parameter regimes of the model: (a) weak interaction regime, (b) strong interaction regime, (c) intermediate interaction regime. *s-up* and *s-down* indicate spin up channel and spin down channel, respectively.

**2.3 Magnetic moment**

Given the three parameters: $\Delta$, $\delta$ and $J$, we are able to determine the energy sequence of *d*-orbitals (also *d*-bands for periodic systems) and thus derive the local magnetic moment. There are three typical parameter regimes. First, ligand splitting is smaller than exchange splitting ($J > \Delta, \delta$). This regime corresponds to the situation that TM interacts only weakly with the carbon skeleton. Hence, we denote this situation as weak interaction (Fig. 2a). In this situation, valence electrons of TM tend to fill in one spin direction with priority, then to fill in the other. The second regime is $J < \Delta, \delta$ i.e. the exchange splitting is small compared to ligand splitting (Fig. 2b). This regime corresponds to the situation that TM-carbon interaction is very strong, and thus is denoted as strong interaction. In this case, valence electrons tend to fill in a single orbital with priority. Consequently, the local magnetic moment is either 0 or 1 $\mu_B$, depending on the number of valence electrons. The third regime is $\Delta > J > \delta \approx 0$ (Fig. 2c). The exchange splitting is of an intermediate value between the ligand splitting $\Delta$ and $\delta$, thus this situation is denoted as intermediate interaction. In this case, electrons first fill the low-lying four *d*-orbitals, then move to the left six *d*-orbitals. The magnetic moments as a function of the number of valence electron are summarized in Table 1.

**Table 1. The magnetic moments via valence electron number under the three parameter regimes. The two numbers in the brackets indicate the *d*-orbital occupation number in each spin channel.**

| Val-electron→ | n=5 | n=6 | n=7 | n=8 | n=9 | n=10 |
|---|---|---|---|---|---|---|
| Weak | $5\mu_B(5,0)$ | $4\mu_B(5,1)$ | $3\mu_B(5,2)$ | $2\mu_B(5,3)$ | $1\mu_B(5,4)$ | $0\mu_B(5,5)$ |
| Strong | $1\mu_B(3,2)$ | $0\mu_B(3,3)$ | $1\mu_B(4,3)$ | $0\mu_B(4,4)$ | $1\mu_B(5,4)$ | $0\mu_B(5,5)$ |
| Intermediate | $1\mu_B(3,2)$ | $2\mu_B(4,2)$ | $3\mu_B(5,2)$ | $2\mu_B(5,3)$ | $1\mu_B(5,4)$ | $0\mu_B(5,5)$ |

## 2.4 Gap of the band structure

Due to the constraints imposed by symmetry, the energy of impurity bands at the Γ point of the Brillouin zone (BZ) follows the same splitting mode as the orbital splitting. For instance, if $d_{xz}, d_{yz}$ orbitals are degenerate under a certain symmetry, the impurity bands corresponding to $d_{xz}, d_{yz}$ would converge at the Γ point. The energy degeneracy at the Γ point is protected by symmetry, it determines the appearance or disappearance of band gaps in TM-carbon systems.

Consider a system located in the intermediate interaction regime ($\Delta > J > \delta \approx 0$). In this case, the impurity bands $d_{xz}, d_{yz}$ and $d_{z^2}$ form a triple band and have a three-fold energy degeneracy (3-DP) at the Γ point; $d_{xy} d_{x^2-y^2}$ form a double band and have a two-fold degeneracy point (2-DP) at the Γ point. The triple band and the double band are separated by an energy interval of $\Delta$. If in one spin channel the *d*-orbital filling number $n_d$ is 2, the two low-lying bands $d_{xy} d_{x^2-y^2}$ are filled and the Fermi level crosses the gap between $d_{xy} d_{x^2-y^2}$ and $d_{z^2} d_{xz}, d_{yz}$ (Fig. 3a). Consequently, the inherent gap of the carbon skeleton is exposed and the spin channel is non-conducting. If $n_d = 3$, the added electron fills one of the triple bands ($d_{xz}, d_{yz}, d_{z^2}$); then the Fermi level cuts through the triple band (Fig. 3b). The three bands join together at the Γ point, hence this spin channel is metallic. A similar analysis is applicable to the situation of $n_d = 4$ (Fig. 3c). However, when $n_d = 5$ (Fig. 3d), the inherent gap of the carbon skeleton is exposed again. We may therefore conclude that for intermediate interaction the spin channel is metallic if $n_d = 1, 3, 4$ and semiconducting if $n_d = 2, 5$.

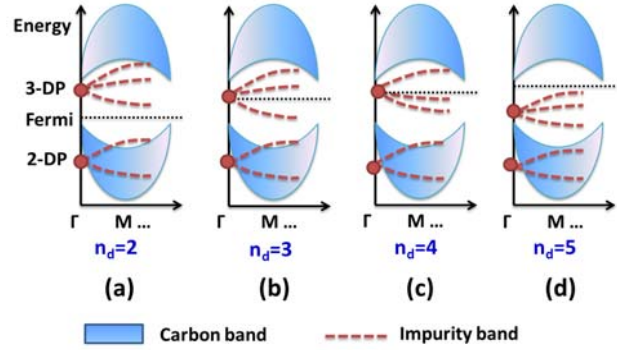

**Figure 3.** Schematic picture of band structures of TM-carbon systems, showing the relation between band gap and the *d*-electron occupation number in one spin channel. (a) When $n_d = 2$, the Fermi level crosses between the "tri-bands" and "di-bands", thus the band gap of carbon structure is exposed; (b) when $n_d = 3$, the Fermi level cuts though the tri-bands and no gap appears due to the degeneracy at Γ point; (c) when $n_d = 4$, it is similar to the case of $n_d = 3$ and there is no gap; (d) when $n_d = 5$, all *d*-bands are underneath the Fermi level and the carbon gap is exposed again.

Consider, for example, Mn doped into a carbon matrix and the ensuing system located in the intermediate interaction regime. Mn has seven valence electrons ($3d^5 4s^2$), Then the *d*-orbital occupation number is 5 in the majority spin and 2 in minority spin (Table 1). Hence there are gaps in both spin channels and the whole system is a semiconductor. For a HM material, we have only to increase the number of electrons by one valence electron. Thus, changing to Fe, which has eight valence electrons ($3d^6 4s^2$), the occupation number in the majority spin is unchanged, while in minority spin it increases from 2 to 3. Consequently, the minority spin becomes a metallic channel and the whole system shows HM. As an alternative, we can decrease the valence electrons by doping Cr ($3d^5 4s^1$). In this case, the majority spin has four electrons and becomes metallic, while that in minority spin has 2 electrons and remains semiconducting; thus the whole system shows HM. It should be noted that for Fe the gap of HM appears in the majority spin, while for Cr it appears in minority spin. Therefore, by varying the doping species one can not only control HM properties but also control the appearance of the gap in the majority or minority spin. Apart from the intermediate interaction regime, we can apply a similar analysis to other systems belonging to strong and weak regimes with similar results.

## 3. First principle calculation

### 3.1 Calculation details

In this part, we employ first-principle methods to calculate a concrete system: TM-graphyne. Spin polarized calculations are performed within first-principles density functional theory (DFT) under the local density approximation (LDA)[27] implemented in the Vienna Ab-initio Simulation Package (VASP)[28, 29]. The projector augmented wave (PAW) method was employed to describe the valence-core interaction.[30, 31] The electronic wave functions are expanded in a plane-wave basis with a kinetic energy cutoff of 400 eV. A vacuum layer of 15 Å is applied in the direction perpendicular to the graphyne plane. An 8 x 8 x 1 Γ-centered Brillouin-zone sampling and a Gaussian smearing with a width of σ = 0.05 eV is used in the electronic structure calculations. All atoms in the supercell were relaxed until residual forces on each atom were less than 0.01eV/Å. The TM-doped graphyne is modeled by a setup of one TM atom in a 2 x 2 graphyne supercell, which corresponds to a concentration of one TM atom per 48 C atoms (Fig. 4). The low concentration removes the direct TM-TM coupling, and the analysis then is focussed on the TM-carbon interaction. Three possible doping sites, T, H and B (green triangles in Fig. 4), are considered. The most stable doping site is at the T site (in the graphyne plane). The adsorption energy is 3~7 eV, which is much greater than that of TM on graphene (~1 eV). We further tested the stability by shifting the TM atoms out of the graphyne plane: the forces on the atom in this case pull the atom back into the the original position. Thus, the in-plane doping at the center of enlarged carbon ring is a stable configuration with $D_{3h}$ symmetry.

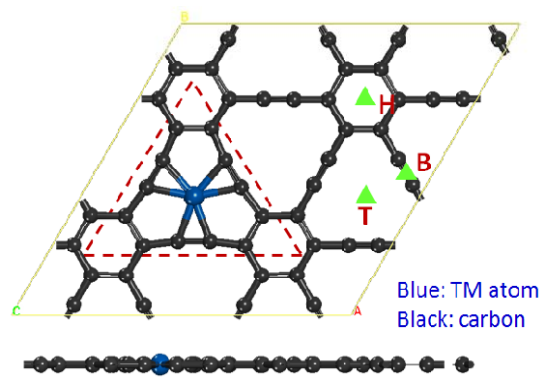

**Figure 4.** Typical configuration of the TM-graphyne system. The parallelogram region denotes the unit cell of the system. Blue and black balls denote TM and C atoms, respectively. The T, H and B mark the possible doping sites we considered. T: on top of the enlarged benzene ring; H: on top of the benzene ring; B: on top of a triple bond.

### 3.2 Local magnetic moment

Most TM-graphyne structures (except for Ni) show magnetism, which mainly arises due to TM $d$-orbitals. The local magnetic moments are listed in Table 2. Valence electrons first occupy the orbitals $d_{xy}$ and $d_{x^2-y^2}$, then occupy the majority spin of $d_{z^2}, d_{xz}, d_{yz}$, and finally occupy the minority spin states of the three orbitals. Therefore, with different valence electron numbers, the systems attain magnetic moments of 2 $\mu_B$, 3 $\mu_B$, 2 $\mu_B$, 1 $\mu_B$ and 0 $\mu_B$ for Cr, Mn, Fe, Co and Ni doped graphyne, respectively. According to the expected magnetic moments in different parameter regimes shown in Table 1, we find that TM-graphyne systems well match the magnetic moments and $d$-orbital filling sequence expected for an intermediate interaction regime ($\Delta > J > \delta \approx 0$).

**Table 2.** Bond length $l$ between the TM atom and the neighboring C atoms, binding energy $E_b$, total magnetic moment per unit cell $M$, magnetic moment of an isolated atom $iso$-$M$ and conducting properties (Cond.). The binding energy $E_b$ is defined as $E_b = E_{TM} + E_{graphyne} - E_{TM+graphyne}$, where $E_{TM}$ and $E_{graphyne}$ are the energy of an isolated TM atom and graphyne, and $E_{TM+graphyne}$ is the total energy of the 2 x 2 graphyne doped by TM.

| Doping TM | $l$ (Å) | $E_b$ (eV) | $M$ ($\mu_B$) | iso-$M$ ($\mu_B$) | Cond. |
|---|---|---|---|---|---|
| Cr | 1.93 | -3.37 | 2.0 | 6.0 | HM |
| Mn | 1.93 | -4.45 | 3.0 | 5.0 | SC |
| Fe | 1.90 | -5.65 | 2.0 | 4.0 | HM |
| Co | 1.91 | -5.45 | 1.0 | 3.0 | HM |
| Ni | 1.92 | -6.87 | 0.0 | 2.0 | SC |

To generalize our model to other TM-carbon systems, we summarize the existing data about the magnetic moments of various doping systems in Fig. 5. The expected magnetic moments in three regimes shown in Table 1 are indicated as dashed lines in Fig. 5. Remarkably, most of TM-carbon systems, from two-dimensional carbon sheets

(like graphene, graphyne) to semi-one-dimensional carbon nanotubes, are located on the line of intermediate interaction (green dashed line). It means that although the values of Δ, $\delta$ and *J* vary from one to anther system, the sequence of their magnitudes is almost an invariant. These data are achieved by independent studies using different calculation parameters (k-sampling, supercell size, exchange-correlation functional, etc.), but they still coincide; thus the model developed can be considered as highly robust and near-universal for doped carbon structures.

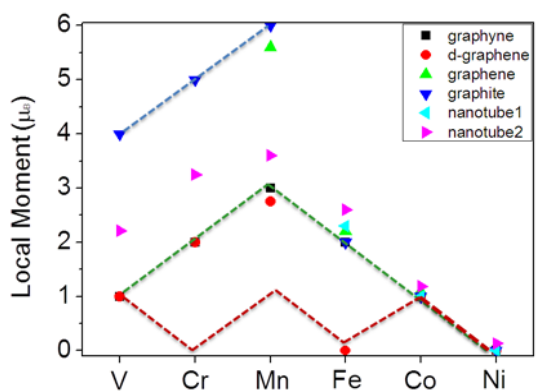

**Figure 5.** The magnetic moments of various TM-carbon systems: TM-graphyne (present work), TM doped into defected graphene (noted as d-graphene)[5], TM doped onto graphene[2], TM doped on graphite surface[32], two kinds of TM doped carbon nanotubes[3, 6]. Blue, green, red dashed lines correspond to the magnetic moments predicted in Table 1 in weak, intermediate, strong interaction regimes, respectively.

Exceptions happen in two special cases. The first is Fe doped graphene with defects, which is located in the strong interaction regime. It is known that defects in graphene induce unpaired electrons which interact strongly with TM (interaction energy ~6-8 eV)[5]. Thus for certain doping species, it is possible to satisfy $\delta > J$. The second case is a V, Cr, Mn doped graphite surface and Mn doped perfect graphene, which are located in the weak interaction regime. Here, the reason is that a perfect graphite surface generally weakly interacts with doping TM (~1-3 eV) [32]. In addition, V, Cr, Mn have a large radius, which further weakens the interaction. This may also explain the fact that magnetic moments of V, Cr, Mn doped nanotubes (nanotube 1,2 in Fig. 5) slightly deviate from the intermediate interaction regime into the weak interaction zone. Generally speaking, strong interactions may appear in case of TM doping into defected carbon sheets; while weak interactions exist in larger TM on perfect surface. But most TM-carbon structures still are firmly in the intermediate interaction regime.

### 3.3 Band gaps

Pure graphyne is a semiconductor with a band gap of about 0.42 eV. The dispersive bands near the Fermi level are due to $p_z$ orbitals of the carbon atoms. Doping TM modifies the carbon band structure by inserting several impurity bands, which can greatly change the electronic properties of the whole system. There are two types of band structures revealed in TM-graphyne systems: HM and semiconducting.

Cr- Fe- Co-graphyne show HM band structures with 100% spin polarization at the Fermi level. Fig. 6a presents the spin-polarized band structure of Cr-graphyne, in which the blue and red circles indicate the contribution from carbon skeleton and TM atoms, respectively. For Cr-graphyne, and in the majority spin channel, there are four occupied $d$-bands (-1.0 ~ 0.0 eV). The Fermi level lies between the two upper $d$-bands, which are degenerate at the $\Gamma$ point. Thus, no gap appears (Fig. 6a). In the minority spin channel (Fig. 6b), there are two occupied $d$-bands (at about -0.6 eV). As there is a large energy gap between the occupied and empty $d$-bands, the inherent band gap of carbon bands is exposed. Fe- and Co-graphyne also show HM, but their gaps appear in the majority spin. For instance, in the majority spin channel of Co-graphyne (Fig. 6c), there are five occupied $d$-bands, all of which lie far below the Fermi level, thus the gap is exposed. In the minority spin channel (Fig. 6d), there are four occupied $d$-bands and the band structure is quite similar to that of majority spin of Cr-graphyne, showing metallic properties.

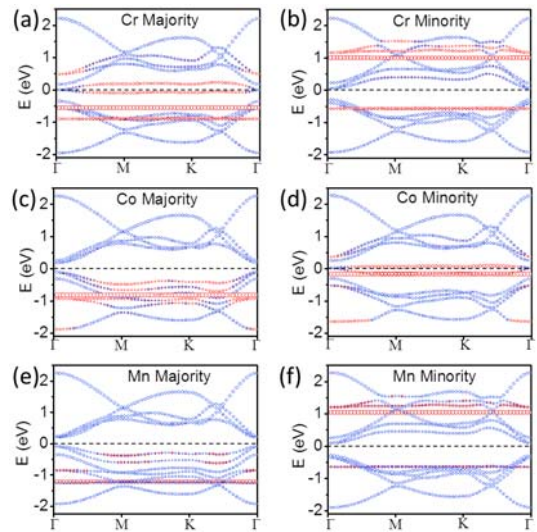

**Figure 6.** Band structures of various TM-graphyne systems. The blue and red circles indicate the contribution from carbon skeleton and TM atoms. (a) (b) band structures of Cr-graphyne in the majority and minority spin channels. (c) (d) Co-graphyne. (e) (f) Mn-graphyne.

Mn- and Ni-graphyne show semi-conducting (SC) properties. In the majority spin channel of Mn-graphyne (Fig. 6e), there are five occupied *d*-bands. In the minority spin channel, Mn-graphyne has two occupied *d*-bands around -0.6 eV (Fig. 6f). Thus the whole system has a magnetic moment of 3 $\mu_B$ per unit cell. For Ni-graphyne, both spin channels have five occupied *d*-bands, thus the whole system is non-magnetic.

The conducting properties of TM-graphyne systems are summarized in Table 2. The conducting properties are determined by the occupation number of electrons in each spin channel. These results are consistent with the predictions of our model.

### 4. Discussion

Rigorously, the model is only applicable to low-concentration cases, because direct TM-TM interactions are neglected. But it still gives a consistent result for higher concentrations and provides a starting point to study more complex cases. Some pioneering work discussed the origin of HM in Heusler alloy[33, 34]. Essentially, these publications rationalized the formation of a gap in HM materials by analyzing valence hybridization due to symmetry. However, a consideration of ligand splitting alone cannot answer the crucial questions why a gap appears in one spin channel and not in the other. In the present work, we retain the basic picture of electrons filling *d*-orbitals, and improve the picture by introducing *J* to describe the exchange splitting. The model thus demonstrates that exchange splitting is essential for understanding the formation of the asymmetric electronic structures. By varying the parameter regime and valence electron number, TM-graphyne shows a large variation of band structures.

### 5. Summary

In summary, we have developed a simplified model to reveal the physical principles for HM, which widely exists in carbon structures. HM is closely related to three common features widely shared by carbon structures: (1) Common splitting mode of *d*-orbitals. We show that the various symmetries in two-dimensional doping systems lead to a common 2-1-2 *d*-orbital splitting mode. Therefore, symmetry becomes a trivial variable and does not explicitly appear in the model. The degeneracy at the Γ point greatly influences the appearance or disappearance of a gap. (2) Most TM-carbon systems lie in the "intermediate interaction" regime. This feature states the fact that the ligand splitting (describing the TM-carbon interaction) is larger than the exchange splitting, but that the two energies are still comparable. The band structure then is due to the balance of the two splitting effects. (3) Carbon based HM materials (TM doped-graphyne, -graphene, etc.) contain conjugated electronic structures, which allow electrons to move freely through the lattice. The three common features are usually concealed by the apparent differences of individual TM-carbon systems.

In addition, we analyzed a realistic HM system: TM atom doped in a graphyne sheet. By changing the doping TM, we are able to tune the valence electron number and hence the conducting properties of each spin channel. Cr-, Fe-, Co-graphyne show HM with magnetic moments of 2 $\mu_B$, 2 $\mu_B$ and 1 $\mu_B$, respectively. The gap of Cr-graphyne shows up in the minority spin channel, while gaps of Fe- and Co-graphyne show up in the majority spin channel. Finally, we show that Mn-graphyne is semiconducting with a magnetic moment of 3 $\mu_B$, and that Ni-graphyne is semiconducting and nonmagnetic. The great flexibility of material properties with different dopant should make transition metal doped graphyne a promising model system for spintronic devices.

## Acknowledgments

This work was financially supported by the MOST (No. 2011CB921702, 2011CB808401), NSFC (51210003), EP/F037783/1 and SSC of China.